\documentclass[a4paper,11pt]{article}
\usepackage{pos}

\usepackage{graphicx,comment}
\usepackage{subcaption}
\usepackage{cleveref}
\usepackage{braket}

\title{Quantum State Preparation for the Schwinger Model}

\author*[a,b]{Giovanni Pederiva}
\author[c,a]{Alexei Bazavov}
\author[a]{Brandon Henke}
\author[a,c]{Leon Hostetler}
\author[a,b]{Dean Lee}
\author[a,c]{Huey-Wen Lin}
\author[a,b]{Andrea Shindler}

\affiliation[a]{Department of Physics and Astronomy, \\
Michigan State University,
East Lansing, 48824, Michigan, USA}
\affiliation[b]{Facility for Rare Isotope Beams,\\
Michigan State University, 
East Lansing, 48824, Michigan, USA}
\affiliation[c]{Department of Computational Mathematics,
Science and Engineering, \\
Michigan State University,
East Lansing, MI 48824, USA}

\emailAdd{pederiva@frib.msu.edu}

\abstract{
It is not possible, using standard lattice techniques in Euclidean space,
to calculate the complete fermionic spectrum of a quantum field theory.
Algorithms running on quantum computers have the potential to access the 
theory with real-time evolution, enabling a direct computation. 
As a testing ground we consider the $1+1$-dimensional Schwinger model with the presence
of a $\theta$ term using a staggered fermions discretization.
We study the convergence properties of two different algorithms ---adiabatic 
evolution and the Quantum Approximate Optimization Algorithm---
with an emphasis on their cost in terms of CNOT gates. This is crucial 
to understand the feasibility of these algorithms, because
calculations on near-term quantum devices depend 
on their rapid convergence. 
We also propose a blocked algorithm that has the first indications of a better scaling 
behavior with the dimensionality of the problem.}

\FullConference{
The 38th International Symposium on Lattice Field Theory, LATTICE2021
26th-30th July, 2021
Zoom/Gather@Massachusetts Institute of Technology
}

\begin{document}
\maketitle

\section{Introduction}
The successful implementation on a noisy intermediate-scale quantum (NISQ) 
device of a gate-based quantum algorithm depends critically on the limited 
and noisy quantum resources. 
The noise is not the only factor to take into account, but also the depth and 
width of a given implementation. In this proceeding we study two different algorithms
for quantum state preparation ---adiabatic evolution~\cite{farhi2000quantum} and the 
Quantum Approximate Optimization Algorithm (QAOA)~\cite{farhi2019quantum}---
on a $1+1$-dimensional quantum field theory, 
the Schwinger model~\cite{schwinger1962gauge}
in the presence of a $\theta$ term~\cite{coleman1975charge}.
The Schwinger model, i.e. Quantum Electrodynamics in $1+1$ dimension, 
has been used in the past as a testing ground for simplified studies of QCD~\cite{Christian:2005yp}
and in the presence of a $\theta$ 
term for testing different computational approaches~\cite{Zache:2018cqq,Funcke:2019zna}.
This is because it enjoys properties such as spontaneous chiral symmetry breaking 
and confinement. After gauge fixing the photons become massive and 
there is a non-vanishing chiral condensate $\left\langle \overline{\psi}\psi\right\rangle$
similarly to QCD.
The presence of a $\theta$ term is particularly interesting because 
direct simulations of QCD with a $\theta$ term in Euclidean space are not possible,
the action being complex.

\section{Model Description}
The Lagrangian of the Schwinger model in $1+1$ dimensions, 
with U($1$) gauge fields $A_{\mu}$, and the $\theta$ term can be written as 
\begin{equation}
    \mathcal{L} = -\frac{1}{4}F_{\mu\nu}F^{\mu\nu}
    + \frac{g\theta}{4\pi}\epsilon_{\mu\nu}F^{\mu\nu} + 
    i\bar{\psi}\gamma^\mu(\partial_{\mu} + igA_{\mu})\psi - m\bar{\psi}\psi\,,
\end{equation}
where the gamma matrices in $1+1$ dimensions are 
$\gamma^0=\sigma^3$, $\gamma^1=i\sigma^2$, $\gamma^5=\gamma^0\gamma^1$,
the field tensor $F_{\mu\nu}$ takes the usual form and $\epsilon_{\mu\nu}$ is a
totally antisymmetric tensor. The parameters of the theory are the gauge coupling $g$,
which in $1+1$ dimensions has the dimension of a mass, 
the fermion mass $m$ and the $\theta$ angle.
Using a $U(1)_A$ rotation, the $\theta$-term can be absorbed into 
the mass term~\cite{Fujikawa:1979ay}. 
If we choose the timelike axial gauge $A_0 = 0$ the Hamiltonian reads~\cite{Hamer:1997dx}
\begin{equation}
    H = \int dx \left[ -i\bar{\psi}\gamma^1 (\partial_1 + igA_1)\psi + 
    m\bar{\psi}e^{i \theta \gamma_5}\psi 
    - \frac{1}{2}\dot{A}_1\dot{A}^1 \right]\,,
\end{equation}
where the electric field has only one component $E=F^{10}=-\dot{A}^1$.
The equation of motion (Gauss's law) 
provides the additional constraint for the remaining gauge degrees of 
freedom 
$\partial_1 E = g \bar\psi\gamma^0 \psi \Rightarrow \partial_1\dot{A}^1 + 
g\bar{\psi}\gamma^0\psi =0$.
Strictly speaking, in a finite volume this is true only with open boundary conditions,
while periodic boundary conditions will still allow a remaining gauge degree of freedom.

\subsection{Discretization and Pauli Hamiltonian}
Following Ref.~\cite{chakraborty2020digital}, in the Hamiltonian formalism 
we keep the time continuous and discretize the spatial dimension on a 1D lattice of $N$ 
sites and lattice spacing $a$ using staggered fermions~\cite{kogut1975hamiltonian,susskind1977lattice}
\begin{equation}
    H = -i\sum_{n=1}^{N-1}\left(\frac{1}{2a} - (-1)^n\frac{m}{2}\sin\theta \right) 
    \left[\chi_n^\dagger e^{-aigA^1(an)}\chi_{n+1} - \text{h.c.} \right] + 
    m\cos\theta\sum_{n=1}^N(-1)^n\chi_n^\dagger \chi_{n} + \frac{g^2a}{2} \sum_{n=1}^{N-1} \Pi^2_n\,,
\end{equation}
where $\Pi_n = -\frac{\dot{A}^1(an)}{g}$ is the rescaled conjugate momentum 
at the lattice point $x=an$.
The fermions have been translated into a pair of one-component spinors 
such that: $\frac{\chi_n}{\sqrt{a}} = \psi_u(an)$ for $n$ even 
and $\frac{\chi_n}{\sqrt{a}} = \psi_d(an)$ for odd $n$. 
In this formulation the Gauss's law becomes~\cite{muschik2017u}:
\begin{equation}
    0 = -(\Pi_n - \Pi_{n-1}) + \chi_n^\dagger \chi_{n} - \frac{1-(-1)^n}{2}\,.
\end{equation}
The $\chi$ field can be transformed into a qubit formulation 
using the well-known Jordan-Wigner transformation~\cite{wigner1928paulische} 
that transforms the fermionic variables into spin variables
\begin{equation}
    \chi_n = \prod_{l<n} i Z_l \frac{X_i - i Y_i}{2}\,,
\end{equation}
where the spin variables are the Pauli matrices located at each lattice 
point, $X_i=\sigma_i^x$, $Y_i=\sigma_i^y$, $Z_i=\sigma_i^z$.
Using open boundary conditions, i.e. fixing the conjugate momentum $\Pi$ at the 
boundary, and solving Gauss's law one obtains
\begin{equation}
    \Pi_n = \Pi_0 + \frac{1}{2}\sum_{l=0}^n\left( Z_l + (-1)^l \right)\,,
\end{equation}
and the value of $\Pi_0$ specifies the boundary conditions.
Removing $\Pi_0$ is equivalent to shifting the $\theta$ angle by 
$2\pi \Pi_0$~\cite{coleman:1976uz},
thus we can safely set $\Pi_0 = 0$.
The parallel transport in the spatial direction
can be reabsorbed in a redefinition of the fermion fields
as an additional phase
$\chi_n \rightarrow \prod_{l<n}\left[e^{-iagA^1(al)}\right]\chi_n$.

The final Hamiltonian, omitting constant terms, can be decomposed as
$H = H_{ZZ} + H_{\pm} + H_{Z}$, where
\begin{align}
    \label{eq:hamiltonian}
    \nonumber
    H_{ZZ} &= \frac{J}{2}\sum_{n=2}^{N-1}\sum_{1\leq k < l \leq n} Z_k Z_l \\
    H_{\pm} &= \frac{1}{2}\sum_{n=1}^{N-1}
    \left(w - (-1)^n\frac{m}{2}\sin\theta \right) 
    \left[X_n X_{n+1} + Y_n Y_{n+1} \right]\\
    \nonumber
    H_{Z} &= m\cos\theta\sum_{n=1}^N (-1)^n Z_n 
    -\frac{J}{2}\sum_{n=1}^{N-1} (n ~\text{mod} ~2) \sum_{l=1}^n Z_l\,,
\end{align}
where, using the same notation as Ref.~\cite{chakraborty2020digital}, we denote
the relevant couplings for the adiabatic evolution as $w=\frac{1}{2a}$ and $J=\frac{g^2}{2a}$.

\section{Adiabatic State Preparation and the Quantum Approximate Optimization Algorithm}
To study the properties of the system on a 
NISQ machine~\cite{Preskill2018quantumcomputingin},
one needs an efficient method to prepare the quantum state. 
The efficiency of a gate-based quantum algorithm depends
on its depth (number of sequential gates) and width (number of qubits).
Besides the optimization at compilation time, 
it is important to optimize the algorithm independently of the hardware used.
In this work, we focus on the depth of two algorithms, trying to 
reduce the number of CNOT operations needed.

\subsection{Adiabatic State Preparation}
Adiabatic State Preparation (ASP)~\cite{farhi2000quantum} is a well-established 
method for quantum state preparation. The basic idea is to first solve a system 
that is simpler than the target one, but for which state preparation is trivial. 
One then slowly (adiabatically) changes the Hamiltonian of the system to reach the target one. 
Following Ref.~\cite{chakraborty2020digital}, we consider as the initial Hamiltonian 
$H_0 = H_{ZZ} + H_Z|_{m\rightarrow m_0, \theta \rightarrow 0}$, which has a ground
state that can be easily identified as the product state of alternating spins up and down. 
The ground state of the Schwinger model is then obtained with an adiabatic Hamiltonian
$H_A(t)$, which interpolates between 
$H_A(t=0) = H_0$ and the Schwinger model, i.e. $H_A(t=T) = H$.

In the simplest case~\cite{chakraborty2020digital} one defines a ``time'' evolution operator 
$U(t) = e^{-iH_A(t)\delta t}$ where $\delta t = \frac{T}{M}$ is the elementary 
step of the adiabatic evolution made of $M$ steps,
and the remaining parts of the Hamiltonian are switched on making the parameters 
$w=\frac{1}{2a}$, $\theta$ and $m$ time dependent
\begin{equation}
        w\rightarrow \frac{t_n}{T} w~~~~\theta\rightarrow \frac{t_n}{T} \theta~~~~~
    m\rightarrow \left(1 - \frac{t_n}{T} \right)m_0 + \frac{t_n}{T} m\,,
    \label{eq:linear}
\end{equation}  
where $t_n = n \frac{T}{M} = n \delta T$, with $n=0,\ldots,M$. 
It is clear that the initial time $t=0$
corresponds to $t_0$ and the final time $T$ corresponds to $t_M$.
We label this linear discretization of the adiabatic evolution $L1$ or $L2$ 
depending on the order of the Trotter product formula we use to evaluate the 
evolution $U(t) = e^{-iH_A(t)\delta t}$.
We have found more efficient discretizations of the adiabatic evolution that we discuss in 
~\Cref{sec:results}.

\subsection{Quantum Approximate Optimization Algorithm}
\label{ssec:qaoa}
The Quantum Approximate Optimization Algorithm (QAOA)~\cite{farhi2019quantum} is a 
quantum optimization algorithm that can also be used for state preparation. 
It has the advantage of having 
a rather shallow depth of $M$ layers and $2M$ total parameters.
Just as ASP, it relies on the existence of a trivially solvable Hamiltonian $H_0$
with eigenstate $\left.|\psi_0\right\rangle$. 
The ansatz for the eigenstate of the target Hamiltonian is then 
\begin{equation}
    \label{eq:ansatz}
    \ket{\psi_M( \vec{\gamma}, \vec{\beta})} = 
        e^{-i\beta_M H_0}e^{-i\gamma_M H}\dots 
        e^{-i\beta_2 H_0}e^{-i\gamma_2 H}
        e^{-i\beta_1 H_0}e^{-i\gamma_1 H} \ket{\psi_0}
\end{equation}
where $\vec{\beta}, \vec{\gamma}$ are $2M$ real coefficients. 
The problem is reduced to finding the optimal values for $\vec{\gamma}^*$ 
and $\vec{\beta}^*$ such that $\ket{\psi_M( \vec{\gamma}^*,
\vec{\beta}^*)}$ is a good approximation of the desired state. 
The optimal parameters are found by solving the variational problem
\begin{equation}
    \bra{\psi_M( \vec{\gamma}, \vec{\beta})}H
    \ket{\psi_M( \vec{\gamma}, \vec{\beta})} \geq E_0,
\end{equation}
with a minimizer of choice. In this work simulated annealing was used
\cite{kirkpatrick1983optimization}, as it is effective for multi-dimensional 
minimization even in the presence of multiple local minima.

\section{Numerical Results}
\label{sec:results}
We summarize here first the results obtained using ASP and QAOA with a 
lattice of $N=4$ points and then present preliminary results for a 
novel blocking procedure up to $N=10$.
\begin{figure}[hbt!]
    \centering
    \begin{subfigure}{0.49\linewidth}
        \centering
        \includegraphics[width=1\linewidth]{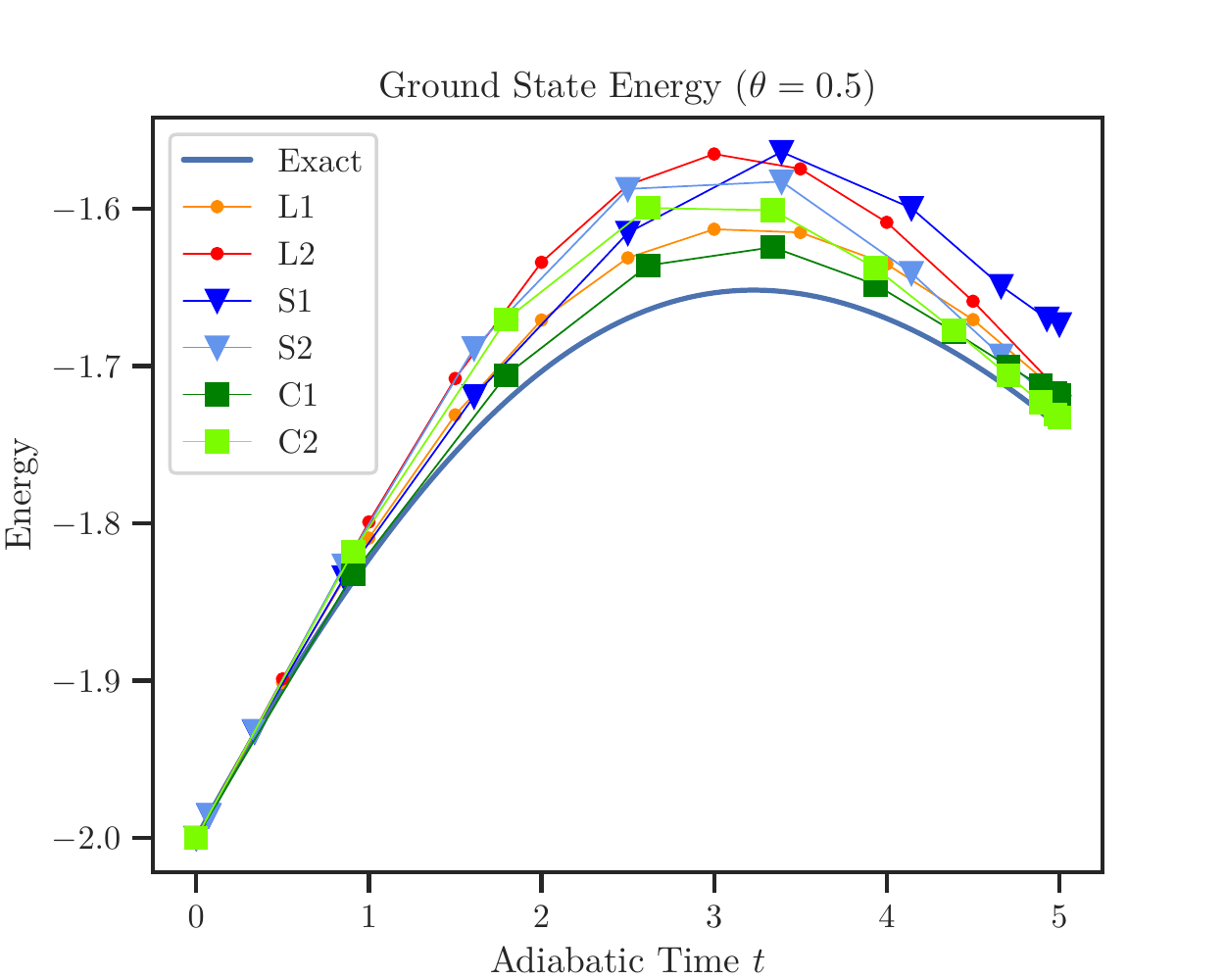}
        \caption{\phantom{b}} \label{fig:1c}
    \end{subfigure}
    \begin{subfigure}{0.49\linewidth}
        \centering
        \includegraphics[width=1\linewidth]{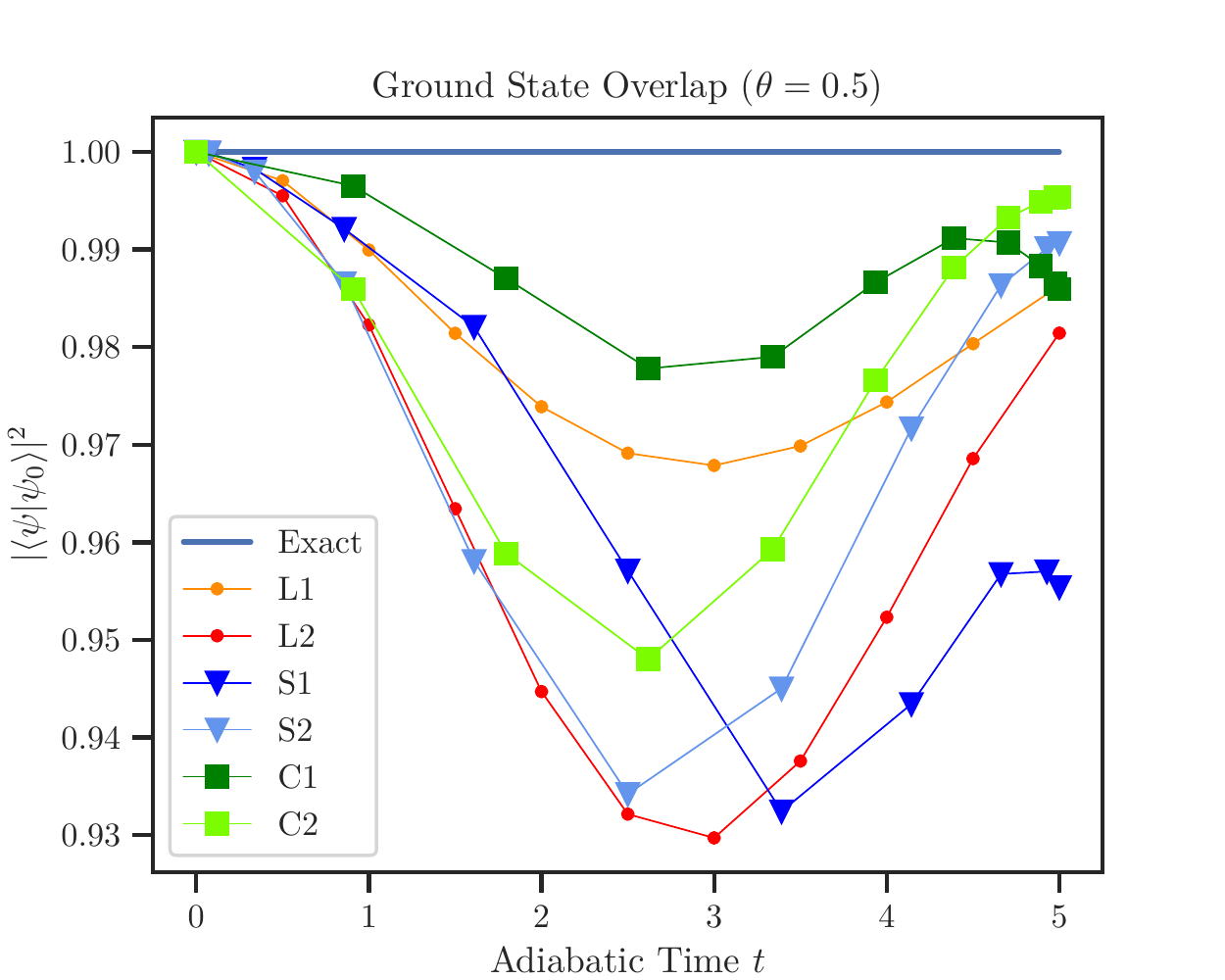}
        \caption{\phantom{b}} \label{fig:1d}    
    \end{subfigure}
    \caption{Simulation of ASP of a $N=4$ system with $J=0.5$, $m_0=0.5$, $m=0$, 
    $w=0.5$, $\theta=0.5$. The different series represent different adiabatic time spacings: 
    linear, $L1$ and $L2$, $S1$ and $S2$ 
    (taking more dense points at the beginning and end of the 
    ASP), and $C1$ and $C2$, (taking more dense points at the end of the evolution), 
    with different Trotterization scheme orders. The simulation is done using 
    the Qiskit package from IBM. Left plot: energy of the system, the exact line 
    represents the results obtained from exact diagonalization of the adiabatic 
    Hamiltonian at every time point. Right plot: overlap between the evolved 
    state and the exact ground state of the adiabatic Hamiltonian obtained from 
    full diagonalization at every time $t$, that is 
    $|\langle\psi(t)|\psi_0(t)\rangle|^2 $. }
    \label{figure:ate}
\end{figure}

\subsection{Results with ASP}
We have simulated an $N=4$ Schwinger model with $g^2=1$, $m_0 = 0.5$, $m=0$, $a=1$ and $\theta=0.5$,
using different discretizations for the adiabatic evolution and different orders
for Trotter's product formula for the time evolution.
%\gio{~\cite{} do we need a citation here? I cannot find any original paper, 
%apparently it is a very old formula by Lie}. 
For the adiabatic evolution, besides the linear discretizations in Eq.~\eqref{eq:linear},
we have considered the following discretizations
\begin{equation}
 t_n = 2 \frac{T}{M}\sum_{k=0}^{n} \sin^2\left(\pi \frac{k}{M}\right)\,,
    \label{eq:sin2}
\end{equation}  
and 
\begin{equation}
    t_n = 2 \frac{T}{M}\sum_{k=0}^{n} \cos^2\left(\pi \frac{k}{2M}\right)\,,
       \label{eq:cos2}
\end{equation}  
which we call respectively $S1$ (or $S2$) and $C1$ (or $C2$) again whether
for the adiabatic time evolution we use a first order or a second order 
Trotter product formula. 
In Fig.~\ref{fig:1c} we show the evolution with ASP of the energy 
of the ground state using ASP with several discretizations of the time evolution. As a 
comparison, we also plot the result for the exact ground state determined for each time value.
We generally observe a very good agreement for the ground state using only $10$ steps
of ASP. We do observe that the $S1$ discretization is the worst at reproducing the ground state,
while the best discretizations are $C1$ and $C2$. 
These discretizations adapt the time interval to be smaller towards the end of the evolution
indicating that it is more relevant to evolve more precisely the state
in the last steps, where the Hamiltonian is closer to the target one, rather than at the 
beginning. These results are summarized in Table~\ref{table:qaoa}.
In Fig.~\ref{fig:1d} we show the results for the overlap with the ground state
as a function of the adiabatic time. The overlap confirms the result of the energy with
the best discretization being $C2$. The results obtained with all the discretizations 
considered here are summarized in Table~\ref{table:qaoa}. 

The drawback of the ASP is that the number of quantum gates required is presently 
too large for any practical purpose. 
Each first-order Trotter step contains $18$ CNOT gates, 
while a second-order step contains $30$. 
This means that for $10$ steps, as in Fig.~\ref{figure:ate}
the total number of CNOTs per qubit is on average $\approx 45$ 
and $\approx 75$ for first and second order Trotterization respectively. 
These numbers (summarized in Table~\ref{table:qaoa}) are too large 
for current quantum hardware. 

\subsection{QAOA Results}
The QAOA method relies on finding the optimal values for the parameters 
$\vec{\beta}, \vec{\gamma}$ introduced in Sec.~\ref{ssec:qaoa}, 
and it has the advantage to allow setting the number of 
steps in the evolution to a relatively small number, provided one can find the optimal 
parameters for such an evolution. For instance, for the same Schwinger model for $N=4$
and same parameter values, we obtained comparable results 
for the energy of the ground state with only $2$ steps, see Table~\ref{table:qaoa}.
\begin{table}
    \centering
    \begin{tabular}[!h]{|c|c|c|c|c|c|}
        \hline
        Method & \# of Steps &\# of CNOT/Qubit & $E_0$ & GS Overlap \\\hline\hline
        ASP $L1$ &  $10$ & $45$ & $-1.7140$ & $0.9827$ \\\hline
        ASP $S1$ &  $10$ & $45$ & $-1.6751$ & $0.9599$ \\\hline
        ASP $C1$ &  $10$ & $45$ & $-1.7144$ & $0.9827$ \\\hline\hline
        ASP $L2$ &  $10$ & $75$ & $-1.7089$ & $0.9729$ \\\hline
        ASP $S2$ &  $10$ & $75$ & $-1.7204$ & $0.9847$ \\\hline
        ASP $C2$ &  $10$ & $75$ & $-1.7260$ & $0.9880$ \\\hline\hline
        QAOA & $2$  & $18$ &      $-1.7353$ & $0.9975$ \\\hline
        QAOA & $3$  & $27$ &      $-1.7357$ & $0.9977$ \\\hline
    \end{tabular}
    \caption{Comparison of final states from ASP and QAOA for the same system 
    as in Fig.~\ref{figure:ate}. For reference, the ground state energy from exact 
    diagonalizaition is $E_0 = -1.7386$. Calculations are performed using the 
    Qiskit software package from IBM~\cite{qiskit}.}
    \label{table:qaoa}
\end{table}

The success of the QAOA algorithm depends on the solution of the optimization
(minimization) problem that needs to be solved either classically or with a 
quantum algorithm. Solving it classically is feasible only for small systems, but
solving it on quantum hardware can be expensive and limited by noise. One option
to improve the scaling of the algorithm would be to use custom optimized 2-qubit
gates~\cite{glaser2015training}. However, since the Hamiltonian is nonlocal 
due to the interaction and boundary term, this cannot be done trivially. 

Another option we propose here is to first solve a blocked system, then stack 
such blocks and use them as a starting point for further optimization. More 
precisely, we solve the $N=4$ with $M=3$ layers of the QAOA algorithm including in the 
layers a modified or ``blocked'' Hamiltonian, $H^B$.
The ``blocked'' Hamiltonian has the same structure of Eqs.~\r{eq:hamiltonian} 
but with only the 1-qubit terms and the 2-qubit ones that connect sites 0 and 1 
(the first block) and the terms that connect qubits 2 and 3 (the second block). 
In this way the four sites are split into two non-interacting blocks and this modified
 Hamiltonian can thus be implemented in a single 2-qubit gate. The complete 
QAOA trial wave function then reads
\begin{equation}
    \ket{\psi_3^B( \vec{\gamma}, \vec{\beta})} = 
        e^{-i\beta_3 H_0}e^{-i\gamma_3 H} 
        e^{-i\beta_2 H_0}e^{-i\gamma_2 H^B}
        e^{-i\beta_1 H_0}e^{-i\gamma_1 H^B} \ket{\psi_0}\,.
\end{equation}
The $M=3$ layers allow correction for any possible ``error'' from the blocked Hamiltonian
adding a layer with the exact Hamiltonian $H$.
The key difference between this ansatz and the $M=3$ equivalent from Eq.~\ref{eq:ansatz}
is that the operator $e^{-i\gamma_n H^B}$ can be implemented more effectively  
as a single gate, greatly reducing the length of the algorithm.

To scale the system to a larger lattice, one can use the same values for 
$\vec{\beta}, \vec{\gamma}$ as found for a smaller system
as a starting guess for the minimizer: this greatly simplifies the variational 
problem. In \Cref{table:qaoa_scaling} we can see that the optimal parameters for
the blocked $N=4$ system give very good results for the ground state of systems 
up to $N=10$, without the need to solve any variational problem at that size.

\begin{table}
    \centering
    \begin{tabular}[!h]{|c|c|c|c|c|}
        \hline
        $N $ & $\#$ CNOT/qubit & $E_0$ & $E_0/E_{\text{Exact}}$ & GS Overlap \\\hline\hline
        $4*$ & $19  $ $(10.5)$ & $-1.7263$ & $0.9931$ & $0.9924$ \\\hline
        $6 $ & $24  $ $(16.6)$ & $-3.4072$ & $0.9926$ & $0.9872$ \\\hline
        $8 $ & $28.5$ $(21)  $ & $-5.6292$ & $0.9926$ & $0.9780$ \\\hline
        $10$ & $32.8$ $(23.2)$ & $-8.3265$ & $0.9930$ & $0.9676$ \\\hline
    \end{tabular}
    \caption{QAOA Blocked with $M=3$ steps, meaning with $2$ steps performed with the 
    ``blocked'' Hamiltonian and $1$ step with the exact Hamiltonian, 
    for the same parameters as in Table~\ref{table:qaoa}. The results for the $N=4$ 
    are obtained after parameter optimization, the results for $N=6,8,10$ have 
    been computed using the same optimal parameters for $N=4$. The number of qubits 
    in parentheses represents the amount of CNOT gates required if optimal custom 
    gates are used.}
    \label{table:qaoa_scaling}
\end{table}

\section{Summary and Outlook}
We found that it is possible to efficiently prepare the ground state of the Schwinger
model with $\theta$-term using both Adiabatic State Preparation and Quantum Approximate
Optimization Algorithm. With the first method, one is limited by the number of 
two-qubit gates in the quantum algorithm, which quickly becomes prohibitively 
large for current quantum hardware. Better adiabatic time discretizations can help 
improve convergence in fewer steps. QAOA on the other hand requires considerably 
fewer gates to achieve comparable results. 

To circumvent the increasing complexity of the optimization problem for QAOA we 
proposed a blocking procedure that produces good candidates for the ground state 
of a system given the optimized parameters of a smaller sized one.

We plan to test these algorithms on real quantum hardware and 
to apply these findings as a starting point for further studies such as 
implementing the Rodeo algorithm~\cite{Choi:2020pdg} to obtain the spectrum of 
the theory.

\section*{Acknowledgements}
This work is supported by the U.S. Department of Energy, Office of Science,
office of Nuclear Physics under grant No. DE-SC0021152.
  
\bibliographystyle{unsrt}
\bibliography{poster}

\end{document}